\documentclass[a4paper,10pt]{article}
\usepackage[utf8x]{inputenc}

\title{Covariant  Graviton Propagator in Anti-de Sitter Spacetime}
\author{Mir Faizal\\ faizal.mir@durham.ac.uk  \\
Department of Mathematics,\\ University of  Durham, Durham, UK  \\
 }

\begin{document}

\maketitle

\begin{abstract}
We construct the graviton propagator in $n$-dimensional  anti-de Sitter spacetime in the most general covariant gauge.  We  then study the behaviour 
of this propagator for  different values  of the gauge parameters. We will show that in any gauge, apart from the Landau gauge, 
the graviton propagator in the AdS spacetime contains a complicated term  involving the derivative of a hypergeometric function which
can not be expressed in terms of elementary functions. We do our calculations in the Euclidean approach. 
\end{abstract}

\section{Introduction}
Anti-de Sitter  (AdS) spacetime is the maximally symmetric  solution to the vacuum Einstein equations with a negative cosmological constant 
\cite{1}. 
It has the topology $S^1 \times R^{n-1}$ and can be viewed as a hyperboloid  in $R^n$. There are closed time like curves in this space. However if we  
unwrap the circle $S^1$, then no closed time-like curves are left. By AdS spacetime we will mean this unwrapped AdS spacetime, in this paper.
 
AdS spacetime has attracted much attention due to AdS/CFT correspondence \cite{2}-\cite{4}. 
This is a correspondence between classical gravity in bulk of AdS and the
 quantum field theory living on its boundary. One important aspect of AdS/CFT correspondence 
is the calculation of correlation functions in the 
type II supergravity 
on AdS in order to study the large N limit of $N=4$ super-conformal Yang-Mills theories \cite{5}-\cite{7}. 
 Bulk to Bulk propagators are required for this purpose \cite{8}-\cite{10}. 
The  graviton propagator in AdS spacetime in a certain Landau gauge is already known \cite{11}.
The graviton propagator for $f(R)$-gravity in four dimensional   AdS spacetime   in a similar 
Landau gauge has  been  recently analysed \cite{1a1}. 
 The Graviton propagator in  AdS spacetime  has also been studied in the  de Donder  and the  
Feynman gauges  
\cite{11a}.
General higher-spin quantum field theory in AdS spacetime has been 
thoroughly investigated in the de Donder gauge \cite{12a}-\cite{12b}.
However, the graviton propagator in AdS spacetime has not been studied in any other gauge. 
In this paper we shall derive the graviton propagator in AdS spacetime in the most general 
covariant gauge. It may be noted that the covariant graviton propagator
 is already know for the de Sitter spacetime \cite{12}-\cite{13}. 

We shall work in the Euclidean approach of Allen and  Jacobson \cite{14}. Thus we shall compute
 the Green's function on the $n$-dimensional hyperboloid and this will become the 
Feynman propagator in the AdS 
spacetime upon analytic continuation. So we will take the Euclidean vacuum \cite{15}
 as the vacuum state for doing our calculations.
We take the boundary condition as the  standard boundary condition in AdS spacetime,
 namely that the fastest possible falloff at the boundary. 
Our propagator   consists of three sectors namely the scalar, the vector and the tensor sectors. 
The tensor sector is the same as that in the work of  D'Hoker,  Freedman, 
Mathur,  Matusis and Rastelli \cite{11}.
  The vector sector can be obtained by a trivial modification of the 
calculations performed for the covariant graviton propagator in de Sitter spacetime \cite{13}. 
So we  really need to generalize only the scalar sector of the graviton propagator in the AdS 
spacetime here. 

\section{The field equation and the Green function}

In perturbative quantum gravity one writes the full metric in terms of a fixed background metric and small perturbations around it.
 We have denoted the full metric as $g^{(f)}_{ab} $ to distinguish it from the fixed  background metric $g_{ab}$. 
We also denote the  small perturbation around the fixed background metric as $h_{ab}$. So we can now write, 
\begin{equation}
 g^{(f)}_{ab} = g_{ab} + h_{ab}, 
\end{equation}
This small perturbation is regarded as a field that is to be quantized. 
 The covariant derivative  along with the raising and lowering of indices is   with  respect 
to the background metric. 

Thus, the action for perturbative quantum gravity in AdS spacetime can be written as 
\begin{equation}
 S = \int d^n x \sqrt{-g} [\mathcal{L}_{\rm{lin}} + \mathcal{L}_{\rm{int}}],
\end{equation}
where $\mathcal{L}_{\rm{lin}}$ is the free part of the Lagrangian  which is quadratic in the field variable and $\mathcal{L}_{\rm{int}}$ is part 
related to interactions.  
Here we are interested only in the free theory so we will take $\mathcal{L}_{\rm{lin}}$ as our Lagrangian. This Lagrangian  is invariant under the 
following gauge transformations, 
\begin{equation}
\delta_\Lambda h_{ab} = \nabla_a \Lambda_b + \nabla_b \Lambda_a.
\end{equation}
We have to break this gauge invariance of the theory before quantizing it.
For this purpose we add the following gauge-fixing term to the original classical  Lagrangian:
\begin{equation}
 \mathcal{L}_{\rm{gf}} = \frac{1}{2 \alpha} \left( \nabla_c h^{cb} - \frac{1+ \beta}{\beta} \nabla^b h\right)
\left( \nabla^a h_{ab} - \frac{1+ \beta}{\beta} \nabla_b h\right),
\end{equation}
where $\beta$ is an non-zero arbitrary finite number. 

To find the Feynman propagator, we have to first find the Euclidean Green's function 
for $\mathcal{L} = \mathcal{L}_{\rm{lin}} + \mathcal{L}_{\rm{gf}} $ and then 
analytically continue it back to  AdS.
The Euler-Lagrange field equation for this total Lagrangian  is given by 
\begin{equation}
 \frac{\partial \mathcal{L}}{\partial h_{cd}} - \nabla_a \frac{\partial \mathcal{L}}{\partial \nabla_a h_{cd}} = 0. 
\end{equation}
 This equation can be written as: 
\begin{eqnarray}
{L_{ab}}^{cd}h_{cd} &= & -\frac{1}{2}\nabla_c \nabla^c h_{ab} + \left( \frac{1}{2}-\frac{1}{2\alpha}\right)
(\nabla_a\nabla_c{h^c}_b + \nabla_b\nabla_c{h^c}_a) \nonumber \\
&& - \left(\frac{1}{2}-\frac{\beta+1}{\alpha\beta}\right)\nabla_a\nabla_b h
- \left[ \frac{(\beta+1)^2}{\alpha\beta^2} - \frac{1}{2}\right]g_{ab}\nabla_c\nabla^c h \nonumber \\
&&  -\frac{1}{2}g_{ab}\left[ 1- \frac{2(1+\beta)}{\alpha\beta}\right]\nabla_c\nabla_d h^{cd}
- R^{-2} h_{ab} - \frac{n-3}{2}R^{-2} g_{ab} h\nonumber  \\  &=& 0. \label{w}
\end{eqnarray}
Here $R$ is the radius of the hyperbolic space from which AdS spacetime is obtained. 

Now we  split the field $h_{ab}$ as follows:
\begin{equation}
h_{bc} = A_{bc}+ B_{bc}  + C_{bc},
\end{equation}
where  $C_{bc}$, $A_{ab}$ and $B_{ab}$ are the   scalar, vector and tensor sectors respectively. 
We  decompose the scalar part into a complete set of modes $C_{ab}^{\lambda\sigma}$  
 constructed from $\phi^{\lambda \sigma}$, where
\begin{equation}
\nabla^2\phi^{\lambda\sigma} = -\lambda\phi^{\lambda \sigma},\label{j}
\end{equation}
and $\sigma$ represents all the other labels. The modes $C_{ab}^{\lambda\sigma}$ can be written as a sum 
of a traceless part $W_{ab}^{\lambda\sigma}$ and the trace $X_{ab}^{\lambda\sigma}$ \cite{16},
\begin{equation}
 C_{ab}^{\lambda \sigma} = X_{ab}^{\lambda \sigma} + W_{ab}^{\lambda \sigma},
\end{equation}
where
\begin{eqnarray}
X_{ab}^{\lambda \sigma} & = & \frac{1}{\sqrt{n}}g_{ab}\phi^{\lambda \sigma},\\
W_{ab}^{\lambda \sigma} & = & \sqrt{\frac{n}{n-1}}\frac{1}{\sqrt{\lambda(\lambda+nR^{-2})}}
\left(\nabla_a\nabla_b + \frac{\lambda}{n}g_{ab}\right)\phi^{\lambda \sigma}.
\end{eqnarray}
Here the factor $1/\sqrt{n}$ and $\sqrt{n}/ \sqrt{(n-1)\lambda + nR^{-2}}$ are
 the normalization factors. The trace of $W^{\lambda \sigma}_{ab}$ obviously vanishes.  
The vector part  $A_{ab}$ is defined  by 
\begin{equation}
 A_{ab}= \nabla_a A_b + \nabla_b A_a, 
\end{equation}
where the divergence of $A_a$ vanishes, $\nabla^a A_a =0$, and 
 this intern implies that the trace of $A_{ab}$ also vanishes, $g^{ab}A_{ab} =0$.
The vector part can also be decomposed using a
 complete set of modes $ A^{\lambda\sigma}_{ab}$ constructed from vector modes,
 $ A^{\lambda\sigma}_{ab} 
= \nabla_a A^{\lambda\sigma}_b + \nabla_b A_a^{\lambda\sigma}$, where \cite{16}
\begin{eqnarray}   
\nabla^a A_a^{\lambda\sigma} &=&0, \nonumber \\
\nabla^2 A_a^{\lambda\sigma}&=& (-\lambda +R^{-2}) A_a^{\lambda\sigma}. \label{vm}
\end{eqnarray} 
Both the trace and divergence of the tensor part $B_{ab}$ vanishes, $\nabla^a B_{ab} = g^{ab}B_{ab} =0$. 
Thus, the tensor part can be decomposed using a complete set of tensor modes $B_{ab}^{\lambda\sigma}$, 
where \cite{16}
\begin{eqnarray}
\nabla^a B_{ab}^{\lambda\sigma} &=& 0, \nonumber \\
g^{ab}  B_{ab}^{\lambda\sigma} &=& 0, \nonumber \\
\nabla^2 B_{ab}^{\lambda\sigma} &=& (-\lambda +2R^{-2})B_{ab}^{\lambda\sigma}.
\end{eqnarray}
Furthermore, the Green's function is  given by 
\begin{equation}
 L_{cd}^{ab} G_{aba'b'} (x,x') = \delta_{cda'b'}(x,x'),
\end{equation}
where for any smooth function $f^{cd}(x)$, we have 
\begin{equation}
 \int d^n x \sqrt{g} \delta_{cda'b'}(x,x')f^{cd}(x) = f_{a'b'}(x').
\end{equation}
Now we can decompose this delta function  and the Green's function into scalar, vector and tensor modes as follows:
\begin{eqnarray}
  \delta_{cda'b'}(x,x') &=&  \delta^{(C)}_{cda'b'}(x,x') + \delta^{(A)}_{cda'b'}(x,x') + \delta^{(B)}_{cda'b'}(x,x'),\\
   G_{cda'b'}(x,x') &=&  G^{(C)}_{cda'b'}(x,x') + G^{(A)}_{cda'b'}(x,x') + G^{(B)}_{cda'b'}(x,x').
\end{eqnarray}
where 
\begin{eqnarray}
 \delta^{(C)}_{cda'b'}(x,x') &=& \sum_\lambda \sum_{\sigma} C^{\lambda \sigma}_{cd}(x)
C^{*\lambda \sigma}_{a'b'}(x'),\\
 \delta^{(A)}_{cda'b'}(x,x') &=& \sum_\lambda \sum_{\sigma}A^{\lambda \sigma}_{cd}(x)
A^{*\lambda \sigma}_{a'b'}(x'),\\
 \delta^{(B)}_{cda'b'}(x,x') &=& \sum_\lambda \sum_{\sigma} B^{\lambda \sigma}_{cd}(x)
B^{*\lambda \sigma}_{a'b'}(x'),
\end{eqnarray}
and 
\begin{eqnarray}
 G^{(C)}_{cda'b'}(x,x') &=& \sum_\lambda \sum_{\sigma} c^{\lambda}_1 C^{\lambda \sigma}_{cd}(x)C^{*\lambda \sigma}_{a'b'}(x'),\\
G^{(A)}_{cda'b'}(x,x') &=&  \sum_\lambda \sum_{\sigma} c^{\lambda}_2A^{\lambda \sigma}_{cd}(x)A^{*\lambda \sigma}_{a'b'}(x'),\\
G^{(B)}_{cda'b'}(x,x') &=&  \sum_\lambda \sum_{\sigma} c^{\lambda}_3B^{\lambda \sigma}_{cd}(x)B^{*\lambda \sigma}_{a'b'}(x'),
\end{eqnarray}
here the sum is a shorthand notation and includes integrals as AdS spacetime is non-compact. 
The constants  $c^{\lambda }_1, c^{\lambda }_2, c^{\lambda}_3$ are  determined by 
\begin{eqnarray}
  L^{ab}_{cd} G^{(C)}_{aba'b'} (x,x') &=& \delta^{(C)}_{cda'b'}(x,x'),\\
 L^{ab}_{cd} G^{(A)}_{aba'b'} (x,x') &=& \delta^{(A)}_{cda'b'}(x,x'),\\
 L^{ab}_{cd} G^{(B)}_{aba'b'} (x,x') &=& \delta^{(B)}_{cda'b'}(x,x').
\end{eqnarray}
We discuss the solutions of these equations in the next  sections.
 \section{Scalar Sector} To find the total propagator,
 we first deal with the scalar sector. First recall that the scalar sector was decomposed 
using $C_{ab}^{\lambda\sigma}$, and this was
  expressed in terms of $X^{\lambda \sigma}_{ab}$
 and $W^{\lambda \sigma}_{ab}$, where 
\begin{eqnarray}
X^{\lambda \sigma}_{ab} & = & \frac{1}{\sqrt{n}}g_{ab}\phi^{\lambda \sigma},\\
W^{\lambda \sigma}_{ab} & = & \sqrt{\frac{n}{n-1}}\frac{1}{\sqrt{\lambda(\lambda+nR^{-2})}}
\left(\nabla_a\nabla_b + \frac{\lambda}{n}g_{ab}\right)\phi^{\lambda \sigma}.
\end{eqnarray}
Furthermore, here $\lambda$ satisfies \cite{16},
\begin{equation}
 \lambda \geq \frac{(n-1)^2}{4 R^2}.
\end{equation}
Now we define $K^{\lambda}_{ij}$, $i,j=1,2$, by
\begin{eqnarray}
{L_{ab}}^{cd}X^{\lambda \sigma}_{ab} = K^{\lambda}_{11}X^{\lambda \sigma}_{ab} + K^{\lambda}_{12}W^{\lambda \sigma}_{ab},\\
{L_{ab}}^{cd}W^{\lambda \sigma}_{ab} = K^{\lambda}_{21}W^{\lambda \sigma}_{ab} + K^{\lambda}_{22}W^{\lambda \sigma}_{ab}.
\end{eqnarray}
We find that $K^{\lambda}_{12}=K^{\lambda}_{21}$. We also define $c^{\lambda}_{ij}$, $i,j=1,2$, by 
\begin{equation}
\left( \begin{array}{cc} c^{\lambda}_{11} & c^{\lambda}_{12} \\ c^{\lambda}_{21} &
 c^{\lambda}_{22}\end{array}\right)
= \left( \begin{array}{cc} K^{\lambda}_{11} & K^{\lambda}_{12} \\ K^{\lambda}_{21} & K^{\lambda}_{22}
\end{array}\right)^{-1}.
\end{equation}
We then find  $c^{\lambda}_{12}=c^{\lambda}_{21}$ and
\begin{eqnarray}
c^{\lambda}_{11} & = & \frac{\beta^2}{n}\left\{\frac{(n-2)\alpha-2(n-1)}{n-2}
\frac{1}{\lambda - (n-1)\beta R^{-2}}\right.\nonumber \\
&& \ \ \ \ \ \ \ \ \ \left. 
- \frac{2(n-1)}{n-2}\frac{\left[n+(n-1)\beta - \frac{n-2}{2}\alpha\beta\right]R^{-2}}{(\lambda - (n-1)\beta R^{-2})^2}\right\},\\
\tilde{c}^{\lambda}_{12} & = & 
\frac{2\beta}{n(n-2)} \frac{n+(n-1)\beta - \frac{n-2}{2}\alpha\beta}{(\lambda - (n-1)\beta R^{-2})^2},\\
\tilde{c}^{\lambda}_{22} & = & - \frac{2}{(n-1)(n-2)R^{-4}}\frac{1}{\lambda+nR^{-2}} + \frac{\alpha}{(n-1)^2R^{-4}}\frac{1}{\lambda}\nonumber \\
&& + \frac{2(n-1)-(n-2)\alpha}{(n-2)(n-1)^2 R^{-4}}\frac{1}{\lambda - (n-1)\beta R^{-2}} \nonumber \\
&& - \frac{2\left[n+(n-1)\beta - \frac{n-2}{2}\alpha\beta\right]}{(n-1)(n-2)R^{-2}}\frac{1}{(\lambda - (n-1)\beta R^{-2})^2},
\end{eqnarray}
where
\begin{eqnarray}
\tilde{c}^{\lambda}_{12} & \equiv & \frac{c^{\lambda}_{12}}{\sqrt{(n-1)\lambda(\lambda + nR^{-2})}}, \\
\tilde{c}^{\lambda}_{22} & \equiv & \frac{nc^{\lambda}_{22}}{(n-1)\lambda(\lambda + nR^{-2})}.
\end{eqnarray}
Define, $\Delta_k(x,x')$ to be the scalar propagator with mass $kR^{-2}$ and 
let $\mu(x, x′)$  be the geodesic distance between spacelike separated points 
$x$ and $x'$ in AdS spacetime and  $ z = \cosh^2 (\mu/ 2R)$,
 then the scalar propagator is given by \cite{14}, 
\begin{equation}
 \Delta_k(z) = q_0 z^{-a_0} F[a_0, a_0 -c_0+ 1; a_0 -b_0 +1; z^{-1}],
\end{equation}
with 
\begin{equation}
 q_0 = \frac{\Gamma(a_0)\Gamma(a_0 -c_0+ 1)}{\Gamma(a_0 -b_0 +1)\pi^{n/2}2^n }|R|^{2-n},
\end{equation}
where 
\begin{eqnarray}
 a_0 &=& \frac{1}{2}\left[ (n-1) + \sqrt{(n-1)^2 + 4 k^2}\right], \nonumber \\
 b_0 &=& \frac{1}{2}\left[ (n-1) - \sqrt{(n-1)^2 + 4 k^2}\right], \nonumber \\
 c_0 &=& \frac{1}{2}n.
\end{eqnarray}
Further, let
\begin{equation}
\Delta_k^{(1)}(x,x') \equiv - \frac{1}{R^{-2}}\frac{\partial\ }{\partial k}\Delta_k(x,x').
\end{equation}
Then the scalar sector of the graviton propagator in AdS, in the Euclidean approach is given by
\begin{eqnarray}
G^{(C)}_{aba'b'}(x,x') & = & g_{ab}(x)g_{a'b'}(x')X(x,x')\nonumber \\
&& + g_{ab}(x)\left(\nabla_{a'}\nabla_{b'} - \frac{1}{n}g_{a'b'}(x')\nabla_{c'}\nabla^{c'}\right)Y(x,x') \nonumber \\
&& +  g_{a'b'}(x')\left(\nabla_{a}\nabla_{b} - \frac{1}{n}g_{ab}(x)\nabla_c\nabla^c\right)Y(x,x') \nonumber \\
&& + \left( \nabla_a\nabla_b - \frac{1}{n}g_{ab}\nabla_c\nabla^c\right)\nonumber \\ && \times 
\left( \nabla_{a'}\nabla_{b'} - \frac{1}{n}g_{a'b'}\nabla_{c'}\nabla^{c'}\right)Z(x,x'), \label{scgf}
\end{eqnarray}
where
\begin{eqnarray}
X(x,x') & = & \frac{\beta^2}{n^2}\left\{\frac{(n-2)\alpha-2(n-1)}{n-2}\Delta_{-(n-1)\beta}(x,x')
\right.\nonumber \\
&& \left.
- \frac{2(n-1)}{n-2}\left[n+(n-1)\beta - \frac{n-2}{2}\alpha\beta\right]\right.\nonumber \\
&& \left. \times R^{-2}\Delta_{-(n-1)\beta}^{(1)}(x,x')\right\}, \\
Y(x,x') & = & 
\frac{2\beta}{n(n-2)}\left[n+(n-1)\beta - \frac{n-2}{2}\alpha\beta\right]\Delta^{(1)}_{-(n-1)\beta}(x,x'),\\
Z(x,x') & = & - \frac{2}{(n-1)(n-2)R^{-4}}\Delta_{n}(x,x') + \frac{\alpha}{(n-1)^2R^{-4}}\Delta_0(x,x')\nonumber \\
&& + \frac{2(n-1)-(n-2)\alpha}{(n-2)(n-1)^2 R^{-4}}\Delta_{-(n-1)\beta}(x,x') \nonumber \\
&& - \frac{2\left[n+(n-1)\beta - \frac{n-2}{2}\alpha\beta\right]}{(n-1)(n-2)R^{-2}}\Delta_{-(n-1)\beta}^{(1)}(x,x').
\end{eqnarray}

Now after finding the graviton propagator in the most general gauge, we will discuses certain limits of this propagator. 
In the gauge,  where $\alpha =0$ and $\beta = n/(1-n)$, we have 
\begin{equation}
 X(x,x') = \frac{2}{(n-1)(n-2)} \Delta_n(x,x'),
\end{equation}
and 
\begin{equation}
Y(x,x') =Z(x,x')=0.     
\end{equation}
So we can see that this part of the graviton propagator can be written in a simple form in the gauge when $\alpha =0$ and $\beta = n/(1-n)$. 
This is the gauge chosen in the work of  D'Hoker,  Freedman, 
Mathur,  Matusis and Rastelli \cite{11}.
Furthermore, for $\beta =0$, we have 
\begin{equation}
 X(x,x') = Y(x,x') =0,
\end{equation}
and 
\begin{eqnarray}
 Z(x,x') & = & - \frac{2}{(n-1)(n-2)R^{-4}}\Delta_{n}(x,x') +\nonumber \\ &&\left[ \frac{\alpha}{(n-1)^2 R^{-4}} + 
  \frac{2(n-1)-(n-2)\alpha}{(n-2)(n-1)^2 R^{-4}}\right]\Delta_{0}(x,x') \nonumber \\
&&  - \frac{2n}{(n-1)(n-2)R^{-2}}\Delta_{0}^{(1)}(x,x').
\end{eqnarray}
Finally we take $\alpha =0$ and $\beta = n(n-2)/4(n-1)$. For this value of $\beta$, the scalar propagator $\Delta_{-(n-1)\beta}(x,x')$
 becomes the propagator for the  conformally coupled case. Thus for $\alpha=0$ and $\beta = n(n-2)/4(n-1) $, we have 
 \begin{eqnarray}
X(x,x') & = & \frac{(n-2)}{2(n-1)} \Delta_{m}(x,x') -\frac{n(n-1)(n+2)}{2(n-2)}R^{-2}\Delta^{(1)}_{m}(x,x'), \nonumber \\ 
Y(x,x') & = & \frac{n(n+2)}{2(n-1)}\Delta^{(1)}_{m}(x,x'),\nonumber \\ 
Z(x,x') & = & - \frac{2}{(n-1)(n-2)R^{-4}}\Delta_{n}(x,x')  + \frac{2}{(n-2)(n-1)R^{-4}}\Delta_{m}(x,x') \nonumber \\
&& - \frac{n(n+2)}{2(n-1)(n-2)R^{-2}}\Delta_{m}^{(1)}(x,x'),
\end{eqnarray}
where  $m = n(2-n)/4$. 
\section{Total Propagator}
To calculate the total graviton propagator we first find the Green's function for the vector part.
The Green's function for  the vector part of the graviton propagator in AdS spacetime is obtained by a trivial modification of
 the calculation's done to calculate the Green's function for the vector part of the graviton propagator in de Sitter spacetime
 \cite{13}. However, 
it will turn out that unlike Green's function for the vector part in de Sitter spacetime 
the Green's function for the vector part  in AdS spacetime contains a very complicated term which can not simply be expressed  in terms of elementary functions.

 To do so we first find the equation of motion for the vector part from by substituting 
$h_{ab} = A_{ab}$ in Eq. $(\ref{w})$. Thus the equation of motion for the vector part is given by 
\begin{equation}
 \frac{1}{\alpha} [-\nabla^2 + (n-1)R^{-2}] A_a =0. 
\end{equation}
 Now  decomposing $A_a$  into a 
complete  set of vector modes and using Eq. $(\ref{vm})$ , we get, 
 \begin{equation}
  \frac{1}{\alpha}[\lambda +(n-2)R^{-2}]A^{\lambda\sigma}_a =0. \label{wq}
 \end{equation}
The Green's function for this equation can be obtained by repeating the calculations that were 
 done for obtaining the Green's function of the vector part in  four dimensional de Sitter 
spacetime \cite{13}, 
 for  $n$ dimensional AdS spacetime. Thus the Green's function 
for Eq. $(\ref{wq})$, can be written as  
\begin{equation}
 G_{ab'}(z)  =-P_{ab'}(z) - \frac{1}{4(n-1)^2} \nabla_a \nabla_{b'} \Delta_0(z).
\end{equation}
If $n_a$ and $n_{b'}$ are unit tangent vectors along the geodesics at $x$ and $x'$ respectively, 
then the parallel transport $g_{ab'}$ is defined by $g^{a}_{c'} n_a (x) = - n_{c'}(x')$.  Now in the  Allen and Jacobson formalism
 \cite{14}, the propagator 
$P_{ab'}(z)$ is given by
\begin{equation}
 P_{ab'}(z) = a(z) g_{ab'} + b(z) n_a n_{b'},
\end{equation}
where $a (z)$ and $b(z)$ again only depend on $z$ which in turn only depends on  the geodesic distance between $x$ and $x'$. They are given by 
\begin{eqnarray}
a(z)&=& \left[\frac{1}{n-1} R \sinh(\mu R^{-1}) \frac{d}{d\mu} + \cosh (\mu R^{-1})\right] \gamma(z),\nonumber  \\ 
b(z) &=& \left[\frac{1}{n-1} R \sinh(\mu R^{-1}) \frac{d}{d\mu} + \cosh (\mu R^{-1}) -1 \right] \gamma(z),
\end{eqnarray}
where
\begin{equation}
 \gamma(z) = \left[ \frac{\partial }{\partial m^2}\left[q_1 F[a_1, b_1; c_1;z]\right]\right] _{| m^2 = 2R^{-2}(n-1)},
\end{equation}
  with 
\begin{equation}
 q_1 = \frac{(1-n)\Gamma(a_1)\Gamma(b_1)}{\Gamma(c_1)2^{n+1} \pi^{n/2} m^2} R^{-n}.
\end{equation}
Here $a_1, b_1$ and $c_1$ are given by
\begin{eqnarray}
 a_1 &=& \frac{1}{2}\left[ (n+1) + \sqrt{(n-3)^2 + 4 m^2 R^2}\right], \nonumber \\
 b_1 &=& \frac{1}{2}\left[ (n+1) - \sqrt{(n-3)^2 + 4 m^2 R^2}\right], \nonumber \\
 c_1 &=& \frac{1}{2}n +1.
\end{eqnarray}
Now  by repeating the argument used to calculate the Green's function of the vector part 
in four dimensional de Sitter spacetime
 \cite{13}, for $n$ dimensional AdS 
spacetime, we obtain the final expression for the  Green's function for the vector  part  of the graviton propagator in AdS spacetime,
\begin{eqnarray}
 G^{(A)}_{aba'b'} (x,x')& =& \alpha [\nabla_a \nabla_{a'}G_{bb'}(x,x') + \nabla_b \nabla_{a'}G_{ab'}(x,x') \nonumber \\ && 
+\nabla_a \nabla_{b'}G_{ba'}(x,x') +\nabla_b \nabla_{b'}G_{aa'}(x,x')].\label{vcgf}
\end{eqnarray} 

As the tensor part does not depend on the gauge parameters it will be the same in all gauge's and the full graviton propagator has already been obtained 
in a certain  Landau gauge \cite{11}. Thus the Green's function for tensor part  in the covariant  gauge will be  the same as that 
which was obtained in that Landau gauge,  however we will include it here for completeness. 
So the Green's function for tensor part  in the covariant gauge can be written as,
\begin{equation}
  G^{(B)}_{aba'b'} (x,x')= \sum_{i=1}^5 G_i(\mu) \mathcal{O}^i_{aba'b'}(x,x'), \label{tcgf}
\end{equation}
where $\mathcal{O}^i_{aba'b'}$ are constructed from all possible linear combinations of the metrics and unit normals at $x$ and $x'$ and along with the
 parallel transport. So we have, 
\begin{eqnarray}
 \mathcal{O}^1_{aba'b'} &=& g_{ab}g_{a'b'},\nonumber  \\ 
 \mathcal{O}^2_{aba'b'} &=& n_a n_b n_{a'} n_{b'}, \nonumber \\ 
 \mathcal{O}^3_{aba'b'} &=& g_{ab'}g_{a'b} + g_{a'b}g_{ab'} ,\nonumber \\  
 \mathcal{O}^4_{aba'b'} &=& g_{ab}n_{a'}n_{b'} +n_{a}n_{b}g_{a'b'} , \nonumber\\  
 \mathcal{O}^5_{aba'b'} &=& g_{ab'}n_{a'}n_{b} + g_{a'b}n_{a}n_{b'} \nonumber \\ && + g_{aa'}n_{b}n_{b'} + g_{bb'}n_{a}n_{a'}. 
\end{eqnarray}
The coefficient $G_i(z)$  also only depend on $z$ and 
 have  been explicitly calculated in terms of a single function $g(z)$ as \cite{11}, 
\begin{eqnarray}
G_1 &=&  \frac{1}{n(n-2)} \left( \frac{4 (z-1)^2 z^2}{n+1} g''(z) + 4z(z-1)(2z-1)g'(z) \right.  \nonumber \\ && 
  + (4nz(z-1) + n-2)g(z)\Big), \nonumber \\ 
G_2 &=& -4(z-1)^2 \left( \frac{z^2}{n(n+1)}g''(z) + \frac{2(n+2)z}{n(n+1)} g'(z) +g(z)\right), \nonumber \\
G_3 &=& \frac{1}{2(2-n)} \left( \frac{4(n-1)z^2(z-1)^2}{n(n+1)}g''(z) + \frac{4(n-1)z(z-1)(2z-1)}{n}g'(z) \right. \nonumber \\ && 
   + (4 (n-1)z(z-1) + n-2)g(z) \Big), \nonumber \\ 
G_4 &=& - \frac{4z(z-1)}{n(n-2)}\left( \frac{z(z-1)}{n+1}g''(z) + (2z-1)g'(z) +ng(z) \right), \nonumber \\ 
G_5 &=& -\frac{x-1}{n-2}\left( \frac{2(n-1)z^2(z-1)}{n(n+1)}g''(z) + \frac{z(4(n-1)z -3n +4)}{n}g'(z) \right. \nonumber \\ &&  
+(2(n-1)z -n +2)g(z) \Big), 
\end{eqnarray}
where  $g(z)$ is given by 
\begin{equation}
 g(z) =g_0(z) + g_1(z) + g_2(z) + g_3(z),
\end{equation}
with 
\begin{eqnarray}
 g_0(z) &=& \frac{n(n+1)\Gamma((n-2)/2)} {(n-1) 2^{n +2} \pi^{n/2}} \frac{d}{dz} \left[\frac{\int_0^z d z' F[-n+2, 1; (4-n)/2; z']}{[z(z-1)]^{n/2}}\right], \nonumber \\ 
g_1 (z)&=& -\frac{n(n+1) \Gamma((n+2)/2)}{2^{n+2} \pi^{n/2}(n-2)(n-1)n} \frac{4(n-1)z(z-1) +n}{[z(z-1)]^{(n+2)/2}},\nonumber \\ 
g_3 (z)&=& \frac{n(n+1)\Gamma((n+2)/2)}{(n-2)(n-1)2^{n+2}\pi^{n/2}}\frac{2z-1}{[z(z-1)]^{(n+2)/2}}, \nonumber \\
g_4(z) &=& (-1)^{(n-1)/2} \frac{\Gamma((n+3)/2)n}{(n+2)(n-1)^2 \pi^{(n-1)/2}} \nonumber \\ && 
\times F[2, n+1; (n+4)/2; 1-z]. 
\end{eqnarray}

Thus, the total graviton propagator in AdS spacetime 
 is given by the sum of the scalar, vector and tensor parts of the graviton propagator. 
We can now write the  graviton propagator, in AdS spacetime, in the most general covariant gauge as:
\begin{equation}
 G_{aba'b'} =  G^{(C)}_{aba'b'} (x,x') +  G^{(A)}_{aba'b'} (x,x') +  G^{(B)}_{aba'b'} (x,x'),
\end{equation}
where  the scalar  contribution to the graviton propagator 
$G^{(C)}_{aba'b'} (x,x')$ is given by Eq. $(\ref{scgf})$,
the vector  contribution to the graviton propagator 
$G^{(A)}_{aba'b'} (x,x')$ is given by Eq. $(\ref{vcgf})$, and 
the tensor  contribution to the graviton propagator 
$G^{(B)}_{aba'b'} (x,x')$ is given by Eq. $(\ref{tcgf})$.

\section{Conclusion}
In this paper we have derived the graviton propagator in the most general covariant gauge. This propagator can be used to calculate the correlation functions in 
the Bulk of AdS spacetime, which in turn can be used to study certain aspects of the AdS/CFT correspondence. It may be noted the the graviton 
 propagator in AdS spacetime  has already been derived in a certain Landau gauge \cite{11}, and used to study certain aspects of the AdS/CFT correspondence 
\cite{17}-\cite{21}. 

We have also seen in this paper that in  any gauge, apart from the 
Landau gauge, the graviton propagator in the AdS spacetime has a very complicated form.  This is because the the vector part of the graviton propagator
in the AdS spacetime contains a derivative of the hypergeometric function which can not be simply expressed in terms of elementary  functions.
As the vector part of the graviton propagator is proportional to the gauge parameter $\alpha$, it vanishes in the Landau gauge. It may be noted that 
in de Sitter spacetime, this problem does not arise and it is possible to write the derivative of the hypergeometric function in terms of elementary 
functions, at least in four dimentions \cite{12}-\cite{13}. As in the work of   D'Hoker,  Freedman, 
Mathur,  Matusis and Rastelli \cite{11}, a certain Landau gauge was chosen, the AdS graviton propagator they derived did not contain this complicated term and 
thus could be easily written in terms of elementary functions.  
\begin{flushleft}
\large{\bf Acknowledgement}
\end{flushleft}
I would like to thank  Atsushi Higuchi for  useful discussions.

\end{document}